\documentclass[a4paper,12pt]{article}
\usepackage{graphicx}
\usepackage{amsmath}
\usepackage{hyperref}
\usepackage{xcolor}
\usepackage{enumerate}
\title{A Family of Local Deterministic Models for Singlet Quantum
State Correlations}

\author{E. Aldo Arroyo\thanks{aldo.arroyo@ufabc.edu.br},\\
    Centro de Ci\^{e}ncias Naturais e Humanas, Universidade Federal do ABC,\\
    Santo Andr\'{e}, 09210-170 S\~{a}o Paulo, SP, Brazil.}

\date{\today}

\begin{document}
    \maketitle
    \begin{abstract}
This work investigates the implications of relaxing the
measurement independence assumption in Bell's theorem by
introducing a new class of local deterministic models that account
for both particle preparation and measurement settings. Our model
reproduces the quantum mechanical predictions under the assumption
of relaxed measurement independence, demonstrating that the
statistical independence of measurement settings does not
necessarily preclude underlying correlations. Our findings
highlight the nuanced relationship between local determinism and
quantum mechanics, offering new insights into the nature of
quantum correlations and hidden variables.
    \end{abstract}

\tableofcontents

\section{Introduction}

We investigate the quantum singlet state, a prototypical example
of entangled spin-1/2 particles. Consider an idealized experiment
where a singlet state is prepared, and the two entangled particles
are sent to two detectors. Bob's detector is aligned along the
unit vector $x$, while Alice's is along the unit vector $y$.

Upon measurement, Bob and Alice observe that each particle's spin
is either aligned or anti-aligned with the respective vectors $x$
and $y$. The possible outcomes for these spin eigenvalues are $+1$
and $-1$, assuming $\hbar = 2$.

Repeating the experiment yields lists of spin projection values
along $x$ and $y$. Denote Alice's outcomes by $A$ and Bob's by
$B$. While each measurement appears random, with $A$ and $B$
taking values $\pm1$, a correlation emerges from their combined
results. Specifically, the expectation value $\langle AB \rangle$,
as predicted by quantum mechanics, is given by
\begin{equation}
\langle AB \rangle = E(x, y) = -\cos\phi_{xy},
\end{equation}
where $\phi_{xy}$ is the angle between $x$ and $y$.

To account for quantum results while maintaining determinism and
realism, hidden variables were introduced. John Stewart Bell
proposed a set of hidden variables $\lambda$, and under the
assumptions of locality, realism, and measurement independence,
$E(x, y)$ can be expressed as
\begin{equation}
\label{densitya1} E(x, y) = \int d \lambda \, \rho(\lambda) A(x,
\lambda) B(y, \lambda),
\end{equation}
where $A(x, \lambda)$ and $B(y, \lambda)$ are measurement outcomes
as functions of settings $x$ and $y$ and hidden variables
$\lambda$, and $\rho(\lambda)$ is the probability density of
hidden variables.

The challenge is to demonstrate that no $\rho(\lambda)$ can
reproduce the quantum prediction for all $x$ and $y$. Directly
proving this requires testing all possible forms of
$\rho(\lambda)$, which is a complex task. Instead, Bell
inequalities, which must be satisfied by $E(x, y)$, offer a more
practical approach \cite{bell1964einstein}. The Bell inequality is
given by
\begin{align}
\label{densityb7} |E(x, y) - E(x, z)| \leq 1 + E(y, z),
\end{align}
where $x$, $y$, and $z$ are arbitrary unit vectors. It is
straightforward to show that the quantum prediction $E(x, y) = -
\cos\phi_{xy}$ violates this inequality. For example, consider
$x$, $y$, and $z$ such that the angle between $x$ and $y$ is 90
degrees, and the angles between $x$ and $z$ and between $z$ and
$y$ are both 45 degrees. Quantum mechanics predicts $E(x, y) = 0$,
$E(x, z) = E(y, z) = -\frac{1}{\sqrt{2}}$. Plugging these values
into Eq. (\ref{densityb7}), we observe a gross violation of the
inequality.

The original Bell inequality, given in Eq. (\ref{densityb7}), is
not typically used in actual experiments. Due to practical
challenges, the CHSH (Clauser, Horne, Shimony, Holt) inequalities
have become more prevalent. These inequalities generalize Bell's
original formulation and are better suited for experimental
verification, particularly in photon pair experiments
\cite{PhysRevLett.49.91}.

Indeed, most experimental tests of Bell's inequalities have been
carried out using entangled photons rather than spin-1/2
particles. Advances in quantum photonics have enabled greater
control and measurement precision, making it the preferred
platform for testing quantum correlations and Bell-type
inequalities
\cite{PhysRevLett.121.080404,PhysRevLett.115.250402,PhysRevLett.115.250401}.

A critical assumption in deriving Bell-type inequalities is
measurement independence. This assumption is often accepted
without question. Shimony and colleagues demonstrate the
plausibility of this assumption through a theoretical scenario
where coordination among physicists, their assistants, and the
supplier of the experimental apparatus inadvertently leads to the
violation of Bell inequalities within a local and deterministic
framework \cite{shimony1985theory}. Nonetheless, the mere
reasonableness of an assumption does not validate it. Other
conditions in the derivation of Bell inequalities, such as the
absence of signaling or determinism, are also considered
plausible. Hence, it is crucial that all foundational assumptions
of Bell inequalities, including measurement independence, undergo
rigorous scrutiny.

The primary aim of this work is to explore the implications of
relaxing the measurement independence assumption and to introduce
a new family of local deterministic models where hidden variables
influence both particle preparation and measurement settings. Our
model successfully replicates the quantum mechanical predictions,
suggesting that the relationship between local determinism and
quantum mechanics is more nuanced than previously thought.

\section{Relaxing Measurement Independence in Bell's Theorem}
As noted, Bell inequalities rely on key assumptions. We now
explore the consequences of relaxing the measurement independence
assumption, building on the work of \cite{brans1988bell}.

When both particle preparation and measurement settings depend on
hidden variables \(\Lambda\), Bell's formula for the expectation
value \(E(x,y)\) must be modified. We use the uppercase variable
\(\Lambda\) instead of the lowercase \(\lambda\) previously used
to denote hidden variables. While Bell's original \(\lambda\)
represents hidden variables associated solely with the preparation
of the entangled particle pair, it does not account for the
influence of the measurement settings \(x\) and \(y\).

In a more general deterministic framework, it is reasonable to
consider that the measurement settings \(x\) and \(y\) could also
depend on additional hidden variables. To account for this broader
scope, we use \(\Lambda\) to denote a set of hidden variables,
which includes not only those related to the preparation of the
particle pair but also those affecting the measurement settings.

This distinction allows us to examine how these additional
dependencies might modify the predicted correlations and
potentially relax the assumptions underlying Bell's original
formulation. As we proceed, we will specify which components of
\(\Lambda\) are related to particle preparation and which are
associated with the measurement settings, ensuring a clear
separation of these influences in our analysis.

Thus, the updated expression for the expectation value \(E(x, y)\)
is given by
\begin{align}
\label{densitya2} E(x, y) &= \frac{\int d\Lambda \, \rho(\Lambda)
\, A(x, \Lambda) \, B(y, \Lambda) \, w(\Lambda; x, y)}{\int
d\Lambda \, \rho(\Lambda) \, w(\Lambda; x, y)},
\end{align}
where \(w(\Lambda; x, y)\) is a weight function depending on both
\(\Lambda\) and the measurement settings \(x\) and \(y\). This
function ensures that only relevant contributions of \(\Lambda\)
are considered. Specifically:
\begin{itemize}
    \item \textbf{If \(\Lambda\) is related to \(x\) and \(y\):} The weight function \(w(\Lambda; x, y)\) is positive, indicating that these hidden variables contribute to the expectation value.
    \item \textbf{If \(\Lambda\) is unrelated to \(x\) and \(y\):} The weight function \(w(\Lambda; x, y)\) is zero, excluding irrelevant hidden variables.
\end{itemize}

Although \(x\) and \(y\) can be freely selected, the outcomes are
governed by \(\Lambda\). Specifically, for each pair \(x\) and
\(y\), there is a relevant subset of \(\Lambda\), and the weight
function \(w(\Lambda; x, y)\) filters \(\Lambda\) to include only
those values associated with the chosen \(x\) and \(y\).

We redefine:
\begin{align}
\label{densitya3} \tilde{\rho}(\Lambda) = \frac{\rho(\Lambda) \,
w(\Lambda; x, y)}{\int d \Lambda \, \rho(\Lambda) \, w(\Lambda; x,
y)},
\end{align}
where \(\int d \Lambda \, \tilde{\rho}(\Lambda) = 1\). Thus,
\(E(x, y)\) can be expressed as:
\begin{align}
\label{densitya4} E(x, y) = \int d \Lambda \,
\tilde{\rho}(\Lambda) \, A(x, \Lambda) \, B(y, \Lambda).
\end{align}

To highlight the role of \(x\) and \(y\) in determining outcomes
influenced by \(\Lambda\), let \(\Lambda = (\lambda, \alpha,
\beta)\), where \(\alpha\) and \(\beta\) are unit vectors in 3D
space representing the hidden variables affecting \(x\) and \(y\).
Therefore, Eq. (\ref{densitya4}) becomes:
\begin{align}
\label{densitya5} E(x, y) = \int d \lambda \, d \alpha \, d \beta
\, \tilde{\rho}(\lambda, \alpha, \beta) \, A(x, \lambda, \alpha,
\beta) \, B(y, \lambda, \alpha, \beta).
\end{align}

The challenge is to determine if the quantum mechanical
prediction:
\begin{align}
\label{densitya6} E(x, y) = -\cos\phi_{xy},
\end{align}
can be replicated by appropriately choosing \(\tilde{\rho}\) in
Eq. (\ref{densitya5}). This problem will be addressed in the next
section, where we examine a specific class of deterministic
models.

\section{A One-Parameter Family of Deterministic Models}
We now introduce a family of deterministic models characterized by
a single real-valued parameter \(\gamma\). To begin, let us
consider the density
\begin{align}
\label{density1} \tilde{\rho}(\lambda, \alpha, \beta) = f(\alpha
\cdot \lambda, \beta \cdot \lambda) |\alpha \cdot \lambda||\beta
\cdot \lambda| \delta(\alpha - x) \delta(\beta - y),
\end{align}
where \( f \geq 0 \) is a function depending on \(\alpha \cdot
\lambda\) and \(\beta \cdot \lambda\). The inclusion of the
absolute values \(|\alpha \cdot \lambda||\beta \cdot \lambda|\)
facilitates subsequent calculations involving \(A\) and \(B\).
This choice ensures positivity and computational convenience.

In this model, \(\alpha\) and \(\beta\) are associated with
measurement settings, while \(\lambda\) influences the outcomes of
\(A\) and \(B\). Consequently, \(A\) and \(B\) are defined as:
\begin{align}
\label{density3}
A(x, \lambda, \alpha, \beta) &= A(x, \lambda) = \text{sgn}(x \cdot \lambda) = \frac{x \cdot \lambda}{|x \cdot \lambda|}, \\
\label{density4} B(y, \lambda, \alpha, \beta) &= B(y, \lambda) =
-\text{sgn}(y \cdot \lambda) = -\frac{y \cdot \lambda}{|y \cdot
\lambda|}.
\end{align}

Integrating over \(\alpha\) and \(\beta\) in Eq. (\ref{densitya5})
yields the expectation value:
\begin{align}
\label{density5} E(x, y) &= - \int d \lambda \, f(x \cdot \lambda,
y \cdot \lambda) \big(x \cdot \lambda \big) \big(y \cdot \lambda
\big).
\end{align}
The normalization condition \(\int d \lambda \, d \alpha \, d
\beta \, \tilde{\rho}(\lambda, \alpha, \beta) = 1\) implies:
\begin{align}
\label{density7} \int d \lambda \, f(x \cdot \lambda, y \cdot
\lambda) |x \cdot \lambda||y \cdot \lambda| = 1.
\end{align}

We express \(\lambda\) in spherical coordinates as \(\lambda =
(\sin \theta \cos \phi, \sin \theta \sin \phi, \cos \theta)\).
With \(x\) and \(y\) chosen as \(x = (1, 0, 0)\) and \(y = (\cos
\phi_{xy}, \sin \phi_{xy}, 0)\), where \(\phi_{xy}\) is the angle
between \(x\) and \(y\). In the following, we consider \(f\) as a
homogeneous function of degree \(k\). That is, the function \(f\)
satisfies:
\begin{align}
f(\mu u, \mu v) = \mu^k f(u, v),
\end{align}
where \(\mu \geq 0\) is a scaling factor and \(k\) is the degree
of homogeneity. Therefore, we can write:
\begin{align}
f\left(\sin \theta \cos \phi, \sin \theta \cos \left(\phi -
\phi_{xy}\right)\right) = (\sin\theta)^k \, f\left( \cos \phi,
\cos \left(\phi - \phi_{xy}\right)\right).
\end{align}
Substituting this expression for \(f\) into Eqs. (\ref{density5})
and (\ref{density7}), and performing the \(\theta\)-integral, we
obtain:
\begin{align}
\label{density11} E(x,y) &= -\frac{\sqrt{\pi} \, \Gamma
\left(\frac{k}{2}+2\right)}{\Gamma \left(\frac{k+5}{2}\right)}
\int_{0}^{2\pi} d\phi
\,  \cos \phi \cos \left(\phi - \phi_{xy} \right) \, f\left( \cos \phi, \cos \left(\phi - \phi_{xy}\right)\right), \\
\label{density12} 1 &= \frac{\sqrt{\pi} \, \Gamma
\left(\frac{k}{2}+2\right)}{\Gamma \left(\frac{k+5}{2}\right)}
\int_{0}^{2\pi} d\phi \, \big|\cos \phi \cos \left(\phi -
\phi_{xy} \right)\big| \, f\left( \cos \phi, \cos \left(\phi -
\phi_{xy}\right)\right).
\end{align}
The Gamma functions appearing in these equations arise from the
integral \( \int_{0}^{\pi} d\theta \, \sin^{3+k} \theta \). It is
important to note that this integral converges for all values of
\(k > -4\).

Let us consider the function
\begin{align}
\label{density13} f(u,v) = \left\{
\begin{array}{ll}
\frac{c_1}{|uv|}(u^2+v^2)^\gamma & \text{if } \text{sgn}(u) = \text{sgn}(v), \\
\frac{c_2}{|uv|}(u^2+v^2)^\gamma & \text{if } \text{sgn}(u) \neq
\text{sgn}(v).
\end{array}
\right.
\end{align}
This function is homogeneous of degree \(k = 2(\gamma - 1)\).
Since \(k > -4\), we require \(\gamma > -1\). The coefficients
\(c_1\) and \(c_2\) are determined by the normalization condition
given in Eq. (\ref{density12}) and by requiring that Eq.
(\ref{density11}) yields the quantum result \(E(x, y) = -\cos
\phi_{xy}\). These coefficients can thus be expressed in terms of
\(\phi_{xy}\) and \(\gamma\), with \(\gamma\) acting as a free
parameter that directly influences the form of the function
\(f(u,v)\).

For generic values of \(\gamma\) and \(\phi_{xy}\), numerical
methods are typically required to evaluate the integrals involving
\(f\). However, for certain specific values of \(\gamma\) and
\(\phi_{xy}\), the integrals in Eqs. (\ref{density11}) and
(\ref{density12}) can be evaluated analytically, providing
explicit analytical expressions for the coefficients \(c_1\) and
\(c_2\). For example, as we will see later, this will be the case
when \(\gamma = 0\). The integrals involved in computing the
coefficients converge if \(\gamma > -\frac{1}{2}\). In the
subsequent numerical analysis, we will restrict our consideration
to \(\gamma > -\frac{1}{2}\).

\begin{figure}[h]
\centering
\includegraphics[width=6.3in,height=49mm]{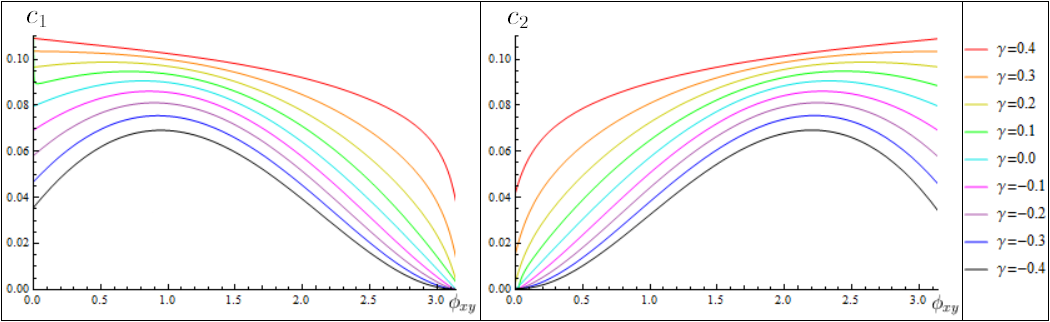}
\caption{Plot of the coefficients \(c_1\) and \(c_2\) as functions
of \(\phi_{xy}\). The curves are colored according to the values
of \( \gamma \): black, blue, purple, magenta, cyan, green,
yellow, orange, and red correspond to \( \gamma = -0.4, -0.3,
-0.2, -0.1, 0, 0.1, 0.2, 0.3, 0.4 \), respectively.} \label{fig1}
\end{figure}

Numerical calculations were performed with \(\phi_{xy}\) in the
range \((0, \pi)\) and \(\gamma\) varying from \(-0.4\) to \(0.4\)
in steps of \(\Delta \gamma = 0.1\). The results for \(c_1\) and
\(c_2\) are shown in Figure \ref{fig1}, where \(c_1\) and \(c_2\)
are plotted against \(\phi_{xy}\). Different colors represent
various values of \(\gamma\), illustrating how the coefficients
vary with \(\phi_{xy}\) and \(\gamma\).

For a given value of \(\gamma\), the graphs reveal a
characteristic: \( c_2(\phi_{xy}) = c_1(\pi - \phi_{xy}) \),
indicating an intrinsic symmetry in the coefficients with respect
to \(\phi_{xy}\). This symmetry will be significant in the
subsequent analysis.

We examine how the densities of the hidden variables vary with
measurement settings \(x\) and \(y\). We quantify differences
between densities \(\rho_{xy}(\lambda)\) and
\(\rho_{x'y'}(\lambda)\) for different settings \((x, y)\) and
\((x', y')\), using a method from \cite{PhysRevLett.105.250404}.

We define the distance between densities \(\rho_{xy}(\lambda)\)
and \(\rho_{x'y'}(\lambda)\) as
\begin{equation}
\label{density14} d(x, y, x', y') = \int d\lambda \, \left|
\rho_{xy}(\lambda) - \rho_{x'y'}(\lambda) \right|,
\end{equation}
and calculate the maximum possible value of \(d\) for all \((x,
y)\) and \((x', y')\). This maximum distance quantifies the
variation in densities \(\rho\) with different measurement
settings.

In our model, the density \(\rho_{xy}(\lambda)\) is given by
\begin{align}
\label{density15} \rho_{xy}(\lambda) = f(x \cdot \lambda, y \cdot
\lambda) \, |x \cdot \lambda| \, |y \cdot \lambda|,
\end{align}
where \(f\) is defined in Eq. (\ref{density13}).

Thus, the distance \(d(x, y, x', y')\) simplifies to:
\begin{align}
\label{density16} d(\phi_{xy}, \phi_{x'y'}) = \frac{\sqrt{\pi} \,
\Gamma (\gamma +1)}{\Gamma \left(\gamma + \frac{3}{2}\right)}
\int_{0}^{2\pi} d\phi \, \big| g(\phi, \phi_{xy}, \gamma) -
g(\phi, \phi_{x'y'}, \gamma) \big|,
\end{align}
where \( g \) is defined in terms of \( f \) as: \( g(\phi,
\phi_{xy}, \gamma) = |\cos \phi \cos (\phi - \phi_{xy})| f(\cos
\phi, \cos (\phi - \phi_{xy})) \). The function \(
d(\phi_{xy},\phi_{x'y'}) \), as defined above, exhibits certain
symmetries with respect to the variables \( \phi_{xy} \) and \(
\phi_{x'y'} \).

Firstly, it is straightforward to observe that \(
d(\phi_{xy},\phi_{x'y'}) = d(\phi_{x'y'},\phi_{xy}) \), indicating
that \( d \) is symmetric with respect to the interchange of \(
\phi_{xy} \) and \( \phi_{x'y'} \).

Secondly, a less obvious symmetry can be demonstrated by
leveraging the previously noted relationship between the
coefficients \( c_1 \) and \( c_2 \), specifically \(
c_2(\phi_{xy}) = c_1(\pi - \phi_{xy}) \). Using this relationship,
it is possible to show that \( d(\phi_{xy},\phi_{x'y'}) = d(\pi -
\phi_{x'y'}, \pi - \phi_{xy}) \).

For a given \( \gamma \), to find the pair \((\phi_{xy},
\phi_{x'y'})\) that maximizes \( d \), we must initially consider
the entire region where \( \phi_{xy} \) and \( \phi_{x'y'} \) can
vary. This region is a square with side length \( \pi \), as both
\( \phi_{xy} \) and \( \phi_{x'y'} \) range from \( 0 \) to \( \pi
\). Therefore, the search for the maximum should cover all pairs
\((\phi_{xy}, \phi_{x'y'})\) within this square.

\begin{figure}[h]
\centering
\includegraphics[width=6.23in,height=67mm]{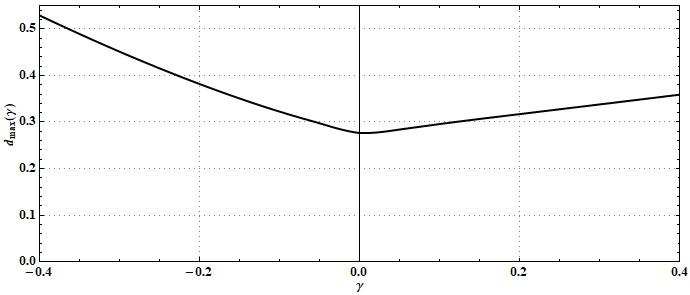}
\caption{The plot shows the maximum values of \( d \) as a
function of the parameter \( \gamma \). The horizontal axis
represents the parameter \( \gamma \), ranging from \(-0.4\) to
\(0.4\), while the vertical axis shows the corresponding values of
\( d_{\text{max}} (\gamma) \).} \label{fig2}
\end{figure}

However, due to the symmetries of \( d(\phi_{xy}, \phi_{x'y'}) \)
discussed previously, we can restrict our search to a smaller
region. Specifically, we focus on the restricted region defined by
\( 0 < \phi_{xy} \leq \pi/2 \) and \( \phi_{xy} \leq \phi_{x'y'}
\leq \pi - \phi_{xy} \). For each calculation, the value of \(
\gamma \) is fixed beforehand, using values within the interval \(
\gamma \in (-0.4, 0.4) \). We numerically compute \( d(\phi_{xy},
\phi_{x'y'}) \) for various configurations of \( \phi_{xy} \) and
\( \phi_{x'y'} \) within the restricted region, identifying the
pairs of angles that yield the maximum value of \( d \). Finally,
we plot these maximum values of \( d \) as a function of \( \gamma
\), denoted as \( d_{\text{max}}(\gamma) \). The results are shown
in Figure \ref{fig2}.

As observed in Figure \ref{fig2}, within the analyzed range of
$\gamma$ from $-0.4$ to $0.4$, the function
$d_{\text{max}}(\gamma)$ attains its minimum value at $\gamma =
0$, indicating that among our family of solutions parameterized by
$\gamma$, there exists a particular member for which
$d_{\text{max}}$ is minimized. For this specific value of $\gamma
= 0$, it is possible to analytically calculate the value of
$d_{\text{max}}$.

First, by setting $\gamma = 0$ (or equivalently, $k = -2$) in Eqs.
(\ref{density11}) and (\ref{density12}), we obtain the following
expressions for the coefficients $c_1$ and $c_2$:
\begin{align}
\label{density18} c_1(\phi_{xy}) = \frac{1+\cos \phi_{xy} }{8 (\pi
-\phi_{xy} )}, \;\;\;\; c_2(\phi_{xy}) = \frac{1-\cos \phi_{xy}}{8
\phi_{xy} }.
\end{align}
This particular case, \(\gamma = 0\), corresponds exactly to the
solution analyzed in \cite{PhysRevLett.105.250404}. Note that,
using these explicit expressions for \(c_1\) and \(c_2\), we can
show that \(c_2(\phi_{xy}) = c_1(\pi - \phi_{xy})\).

It turns out that the maximum value of \( d(\phi_{xy},
\phi_{x'y'}) \) occurs when \( \phi_{xy} + \phi_{x'y'} = \pi \).
Therefore, we will calculate \( d(\phi_{xy}, \phi_{x'y'}) \) under
the assumption that \( \phi_{x'y'} = \pi - \phi_{xy} \). Due to
the symmetry properties of \( d(\phi_{xy}, \phi_{x'y'}) \), we
will use a value of \( \phi_{xy} \) such that \( 0 < \phi_{xy}
\leq \pi/2 \). For \( \gamma = 0 \), using Eq. (\ref{density16}),
we obtain
\begin{align}
\label{density19} d(\phi_{xy},\pi - \phi_{xy}) = \frac{2 \phi_{xy}
+\pi \cos \phi_{xy}-\pi }{\pi -\phi_{xy}}.
\end{align}
Thus, the maximum value of \( d \) will occur at the point \(
\phi_{xy} \) that satisfies the following nonlinear algebraic
equation:
\begin{align}
\label{density20} 1 + (\phi_{xy} - \pi) \sin \phi_{xy} + \cos
\phi_{xy} = 0.
\end{align}

The authors in \cite{PhysRevLett.105.250404} originally reported
that the maximum occurs at \( \phi_{xy} = \pi/4 \). Substituting
this value, \( \phi_{xy} = \pi/4 \), into Eq. (\ref{density19}),
we obtain
\begin{align}
\label{density21} d\left(\frac{\pi}{4}, \frac{3\pi}{4}\right) =
\frac{2}{3} \left(\sqrt{2} - 1\right) \approx 0.276142.
\end{align}
However, we observe that \( \phi_{xy} = \pi/4 \) is not a solution
to Eq. (\ref{density20}). Instead, the correct value of \(
\phi_{xy} \) that maximizes \( d(\phi_{xy}, \pi - \phi_{xy}) \) is
obtained from the numerical solution of the algebraic equation
(\ref{density20}), yielding \( \phi_{xy} \approx 0.81047 \).
Substituting this value into Eq. (\ref{density19}), we obtain \(
d_{\text{max}} \approx 0.276434 \), which is consistent with the
corrected value published in the erratum to
\cite{PhysRevLett.116.219902}.

It is important to emphasize that, unlike the erratum's approach,
which applied numerical maximization directly to find \(
d_{\text{max}} \), our approach derived the maximization condition
analytically via Eqs. (\ref{density19}) and (\ref{density20}), and
numerical calculation was applied solely to solve the nonlinear
algebraic equation (\ref{density20}).

\section{Conclusions}
Through detailed mathematical analysis and numerical calculations,
we have introduced a new family of deterministic models for the
singlet state, parameterized by \(\gamma\), which accurately
reproduces the correlations predicted by quantum mechanics under
the assumption of relaxed measurement independence.

We examined the distance between the densities of the hidden
variables associated with these solutions, as proposed in
\cite{PhysRevLett.105.250404}, to quantify how the densities
differ based on the choice of settings \(x\) and \(y\). For each
value of \(\gamma\) within the range \(-0.4\) to \(0.4\), we
calculated the maximum possible value of \(d\), representing the
greatest difference between the densities. Our analysis revealed
that the smallest of these maximum distances occurs at \(\gamma =
0\), which corresponds to the solution previously considered in
\cite{PhysRevLett.105.250404}.

For the specific case \(\gamma = 0\), our reanalysis identified a
computational error in the calculation of the maximum distance
\(d\). Substituting the corrected value \(\phi \approx 0.81047\)
into Eq. (\ref{density19}) yields \(d_{\text{max}} \approx
0.276434\), which is slightly greater than the previously reported
value of \(d_{\text{max}} \approx 0.276142\). The correct value we
obtained through analytical methods aligns with the value reported
in the erratum published in \cite{PhysRevLett.116.219902}.

Further investigation is needed to determine whether there exists
another family or a specific singlet state model that yields a
distance \(d_{\text{max}}\) satisfying
\begin{align}
\label{conclu1} 0.276142 \leq d_{\text{max}} < 0.276434.
\end{align}
This investigation is particularly important because the lower
bound of \(0.276142\) corresponds to the value defined in the
literature as \(M_{\text{CHSH}}\), which represents the minimum
amount of measurement dependence required to model the maximum
quantum violation of the CHSH inequality
\cite{PhysRevA.84.022102}. Identifying a model that produces a
\(d_{\text{max}}\) within this interval could therefore have
significant implications for understanding the relationship
between measurement dependence and quantum correlations.




\bibliographystyle{ieeetr} 
\bibliography{bibcylib} 
\end{document}